\documentclass[sigconf]{acmart}

\AtBeginDocument{%
  }
    
\usepackage{enumitem} 

\usepackage{multibib}
\newcites{review}{\phantom{x}}

\newcommand{\response}[1]{#1}

                        \setcopyright{none}
                        \acmDOI{}
                        \acmISBN{}

                        \acmConference[LIMITS '26]{12th Workshop on Computing within Limits}{June 23--25, 2026}{Online}
  
\acmISBN{978-1-4503-XXXX-X/2018/06}

\begin{document}

\title{Overview over the first decade of LIMITS}


\author{Maria Emine Nylund}
\orcid{0009-0008-4876-3206}
\email{maria.nylund@sintef.no}
\affiliation{%
  \institution{}
  \city{}
  \country{}
}

\author{Erik Johannes Husom}
\orcid{0000-0002-9325-1604}
\email{erik.johannes.husom@sintef.no}
\affiliation{%
  \institution{SINTEF}
  \city{Oslo}
  \country{Norway}
}

\author{Ophelia Prillard}
\orcid{0000-0003-1744-9720}
\email{Ophelia.Prillard@sintef.no}
\affiliation{%
  \institution{}
  \city{}
  \country{}
}

\renewcommand{\shortauthors}{Nylund et al.}

\begin{abstract} 
  
Computing within limits is a promising field, that follows principles of a) questioning endless growth narrative, b) considering and preparing for models of scarcity and c) reducing energy and material consumption, while considering d) a global spatial scale and e) long time frames. With computing's environmental impact growing and ecological limits becoming increasingly pressing, the LIMITS \response{workshop} has served as a central venue for this community since its inception in 2015, but an overview of the research published there has yet to be described.

This paper addresses this gap by analyzing 160 publications from the LIMITS \response{workshop} in the period 2015 to 2025 to identify its international spread, contributions and developments in relation to field's core concerns, combining programmatic analysis with a manual review.

Our findings indicate that the field has increasingly mentioned degrowth and post-growth, especially in 2024-2025. It has broadened its global perspective, with a growing, but still limited, representation of work beyond the Global North. The majority of papers are positional or observational, while artifact-producing research remains relatively scarce, though solution-oriented output has grown in recent years.

This paper contributes to the LIMITS community by mapping its first decade and current trends to support future research and enhance its global impact.

\end{abstract}




\received{3 April 2026}

\maketitle

\section{Introduction} 

The urgency of reorienting computing research around planetary limits has never been greater. Ripple et al. \cite{ripple2025state} report that of 34 planetary vital signs tracked annually, 22 are currently at record levels, with 2024 as the hottest year on record, while current policies place the planet on a trajectory toward approximately 3.1°C of peak warming by 2100. Seven of \response{nine} globally quantified safe and just planetary boundaries have now been transgressed, and the Greenland and West Antarctic ice sheets may already be passing critical tipping points.
Meanwhile, the information and communication technology (ICT) sector's share of global greenhouse gas emissions continues to grow while increasing computational demands of emerging technologies such as artificial intelligence (AI) boom \cite{bogmans2026power}. 
Against this backdrop, a research community that takes planetary limits as its main principle, rather than an afterthought, is an outright necessity.

Computing Within Limits (LIMITS) emerges from other disciplines such as HCI, psychology, economics, and computer science, but argues that the default assumptions of that ongoing growth are incompatible with planetary boundaries \cite{nardi2018computing}. Although LIMITS shares some concerns with sustainability-oriented fields such as Green IT, it differs from approaches primarily centered on efficiency gains. From a LIMITS perspective, efficiency improvements alone are not enough, as they are frequently offset by the Jevons Paradox: making something more efficient can also make it easier and cheaper, and therefore more widely used.

At the same time, there is a gap in understanding what \textit{computing within limits} means in practice. LIMITS is not only a \response{workshop} series, but a community and a research field centered on computing's place in a world of ecological, material, and social limits. The field has now passed the decade mark and accumulated a substantial publication record, but this record can be difficult for newcomers to navigate. LIMITS papers range from practical, concrete contributions, including software and hardware \cite{rigaud2025zombitron, makonin2022calculating, brain2022solar, sutherland2022strategies, sharmacivil, grinko2021transitions}, to more abstract and reflective work on design patterns, human perception, constraints, and philosophical or political questions \cite{raghavan2015abstraction, joshi2016whose, mcdonald2017political, nystrm2021challenging, houston2022richness}. A review of its publication record can therefore help clarify which topics, methods, and types of contributions have come to dominate the field, and remain comparatively underexplored.

Some LIMITS papers have previously looked at specific parts of the field. \cite{hendry2021transition} examined eight LIMITS papers that address aspects of food, and argued that LIMITS can be seen as a transition system as based on \cite{nardi2018computing}'s statement that LIMITS "contribute to the overall process of transitioning to a future in which the well-being of humans and other species is the primary objective". \cite{silberman2023information} examined changes in the period 2015-2023, noting that environmental problems are primarily political rather than technical, and that technology development can be understood as "politics by other means".

This paper's contribution is a literature review of the full corpus of LIMITS papers from 2015 to 2025, using a combination of programmatic and manual data extraction. \textbf{Our aim} is to investigate how LIMITS has evolved as a venue and a field in terms of publication volume, types of contribution, and to assess how these developments relate to the field's core concerns. We hope to provide both a better overview of what the community has achieved so far, and to offer a clearer basis for the field's future directions.

\section{Method}

This review was done within a critical realist epistemological tradition: we assume that the LIMITS research community has real, observable patterns in publication activity, forms of contribution, and topical emphases, but that our access to those patterns is inevitably affected by the instruments we use, the categories we impose, and the perspectives we bring as researchers. We are ourselves based in Norway, embedded in the Western academic system, without any previous publications at LIMITS. As newcomers, we do not have access to the culture fostered within the community through other channels than the research articles themselves.

\subsection{Programmatic data extraction}

The corpus of this review consists of all retrievable papers published at LIMITS from 2015 to 2025. We identified 162 papers in the proceedings and program of the \response{workshop} throughout its history, but excluded two because only abstracts were available and the full papers could not be found online. 

Most papers were collected through programmatic downloading PDFs from the \response{workshop} websites, while some papers required manual downloading due to bot-protection practices at the ACM Digital Library. The PDFs were batch-processed with GROBID~\cite{GROBID}, an open-source machine-learning library that parses academic PDFs into structured TEI XML, to extract bibliographic metadata. 

For full-text analysis, the body text of each paper was extracted from the GROBID XML, tokenised into alphabetic tokens of at least three characters, and filtered against a curated stopword list combining standard English stopwords and domain-generic academic terms. As a descriptive analysis of geographic orientation, we identified explicit mentions of countries in each paper's body text using a regular-expression gazetteer built from the `pycountry` database, and grouped these mentions against a set of WEIRD (Western, Educated, Industrialised, Rich, Democratic)~\cite{henrichWeirdestPeopleWorld2010}, nations to examine geographic bias in the studies reported. 

After parsing the PDFs with GROBID, all extraction and analysis steps were deterministic and script-based, including regular expressions, tokenisation, frequency counting and rule-based gazetteer matching). We have decided against using methods relying on large language models (LLMs), even though they are increasingly used for performing systematic literature reviews (SLRs)~\cite{scherbakovEmergenceLargeLanguage2025}. While the efficiency and growing capabilities of these models makes it tempting to automate certain steps of the screening and review process, it makes it difficult to reproduce and audit the results. Additionally, it disengages researchers from getting to know the existing research. This design ensures transparency and reproducibility\footnote{The source code for our programmatic data extraction can be found at \href{https://github.com/SINTEF/ComputingWithinLimits-review/}{github.com/SINTEF/ComputingWithinLimits-review}.}. 

We have used LLMs (GPT-5.4, Claude Opus 4.5 and Claude Sonnet 4.6) for writing code used in the analysis and visualization of the results.

\subsection{Manual data extraction}

In addition to the programmatic analysis, we conducted a manual review of the same corpus of 160 papers. The aim was to capture characteristics that are not easy to extract automatically based on metadata analysis and rule-based text processing alone.

An initial pilot study was performed on a small subset of the corpus, to decide on the proper questions to investigate. For the main review, each paper was reviewed on three main aspects: 1) whether the paper was primarily \textit{positional} (perspective, argumentative, discussion), \textit{observational} (case studies, field, studies, analytical, review, descriptive) or \textit{solution-oriented}, 2) whether it presented an artifact (software, hardware, model, framework, etc), and 3) whether it studied users, human behavior or communities. \response{We piloted various ways of categorizing the papers, and landed on the taxonomy of positional/observational/solutional, inspired by a systematic review of Green AI by Verdecchia et al. \cite{verdecchia2023systematicgreenai}, since it gave minimal overlap and ambivalence when labeling papers.} We used a single-reviewer approach, where the papers were divided equally among the authors. Each paper was only reviewed by one person, but ambiguous cases were discussed together to ensure consistent interpretations. Findings from the review also guided the programmatic data extraction as to what categories and trends to look for.

\section{Results}
In this section we present the results of the manual and programmatic data extraction of 160 papers with an average of 3,044 words per paper (std: 1,046).

\subsection{Types of papers} 

Publications show an equal division between \textit{positional}, \textit{observational} or \textit{solution-oriented} types, as seen in Table~\ref{tab:paper_type_summary}.

\begin{table}[h]
\centering
\caption{Overall distribution of paper types.}
\label{tab:paper_type_summary}
\begin{tabular}{lrr}
\toprule
Paper type & Count & Share of all papers (\%) \\
\midrule
Positional & 56 & 35.0 \\
Observation & 54 & 33.8 \\
Solution & 50 & 31.2 \\
\bottomrule
\end{tabular}
\end{table}

\begin{figure}[h]
    \centering
    \includegraphics[width=0.9\linewidth]{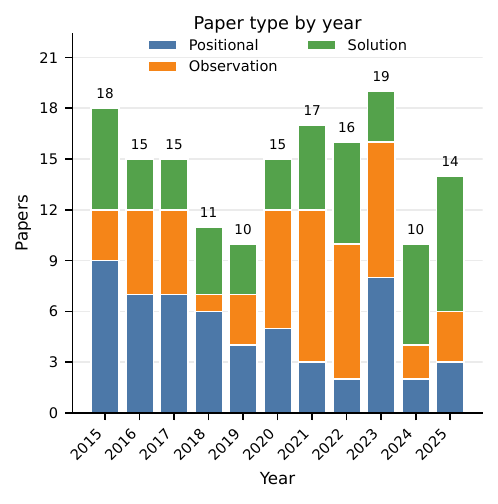}
    \caption{Number of papers per year, separated by whether the papers are positional, observational or solution-oriented}
    \Description{}
    \label{fig:paper_type}
\end{figure}

Figure~\ref{fig:paper_type} presents the distribution of papers per year from 2015 to 2025 where we can observe that there are less positional types of papers in the recent years except for a spike in 2023. There was an increase in observational papers between 2020 and 2023, while solution papers have increased in the last two years.

\begin{figure}[h]
    \centering
\includegraphics[width=0.9\linewidth]{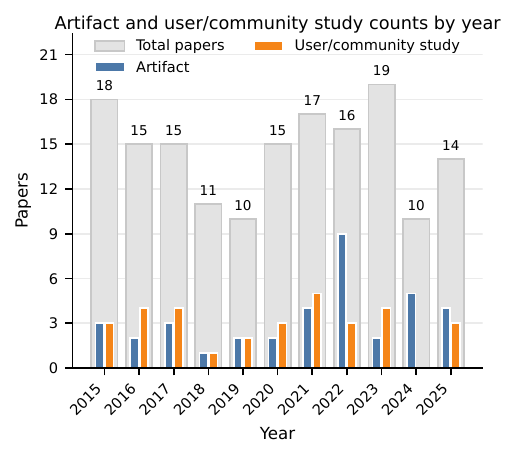}
    \caption{Number of papers that 1) developed an artifact and/or 2) studied user and/or community behavior, compared against the total numbers of paper per year.}
    \label{fig:artifact_user_study}
\end{figure}

\begin{figure}[h]
    \centering
    \includegraphics[width=0.8\linewidth]{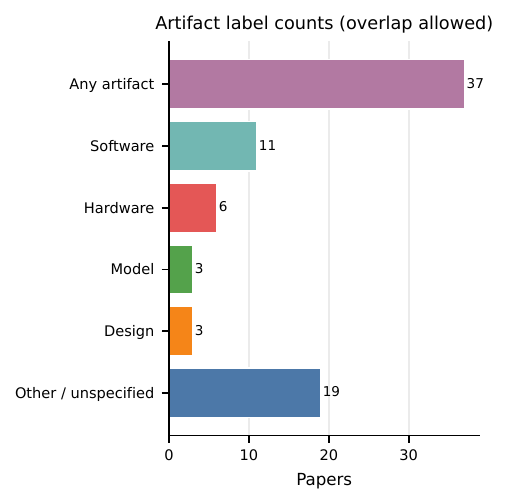}
    \caption{Type of artifacts developed in the research papers with allowed overlap. }
    \label{fig:artifact_artifact_breakdown}
\end{figure}

In Figure~\ref{fig:artifact_user_study}, we can see that papers that studies people (either communities, users or human behaviour) remain consistently low throughout the decade, similarly to development of artifacts. Although year 2022 has resulted in many concrete contributions. Figure~\ref{fig:artifact_artifact_breakdown} shows the breakdown in types of contributions with "other" type being the majority, it included concepts, protocols, guidelines or frameworks for analysis. Software contributions such as \cite{norton2019sage} was the biggest group, where it was often developed together with hardware as in \cite{brain2022solar}. Many papers have analysed existing artifacts, as in \cite{wei2019should} where they analysed 101 design projects, then it was categorised into "observation" category. 

\begin{figure}[h]
    \centering
    \includegraphics[width=1\linewidth]{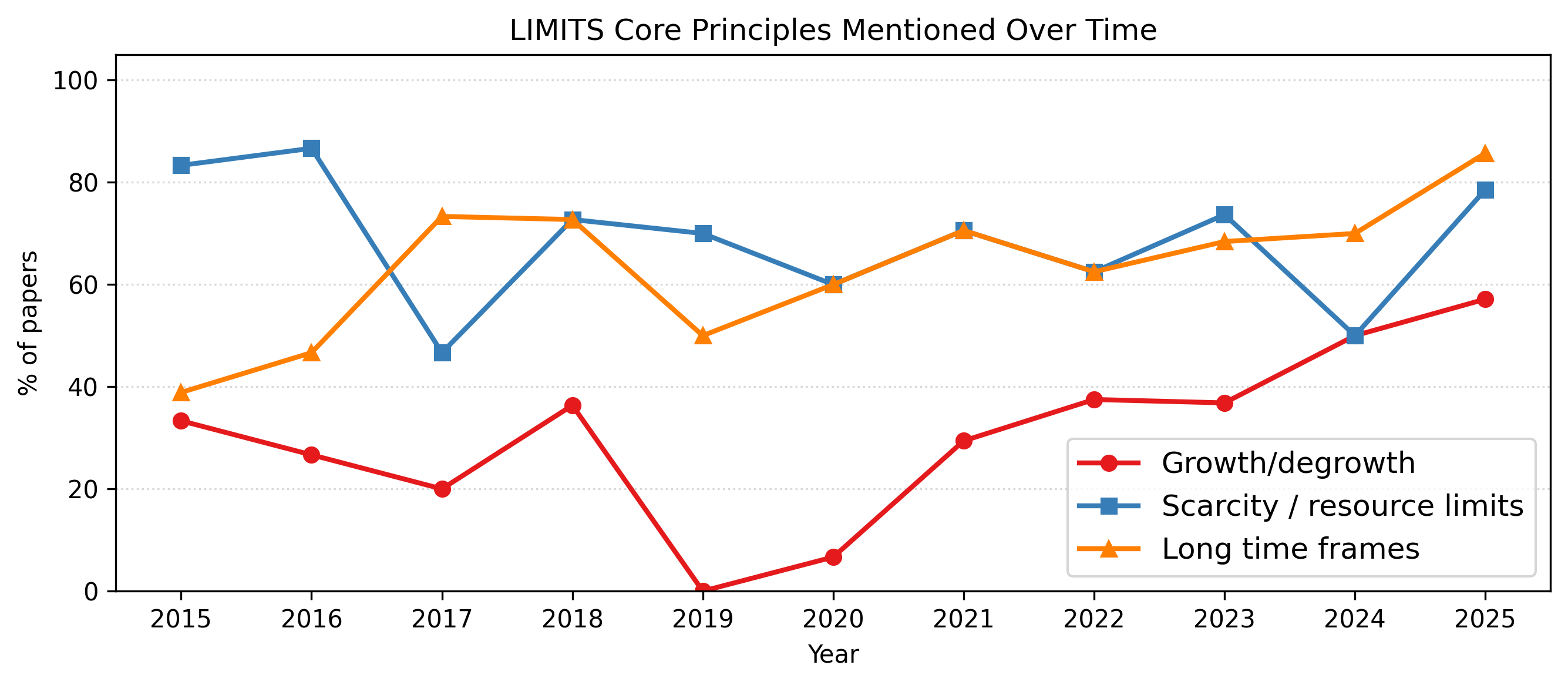}
    \caption{Percentage of articles per year mentioning words related to the main principles. A paper is counted for a principle if its body text contains characteristic terminology; a single paper may score on multiple principles.}
    \label{fig:core_principles}
    \Description{}
\end{figure}

\subsection{Principles} 

To gain an understanding of how the LIMITS principles from Nardi et al.\cite{nardi2018computing} are represented in the publications, we checked whether characteristic terms for each principle appear in the body text of the paper. For "questioning the growth narrative" matching we check for words such as degrowth, post-growth, limits-to-growth framing, "scarcity and resource limits" (e.g. planetary boundaries, resource depletion, resilience); and "long time frames" (e.g. intergenerational equity, multi-decadal horizons, references to 2050 or 2100). 

Figure~\ref{fig:core_principles} shows that "scarcity and resource limits" is the most consistently present principle, appearing in 2/3 of papers across the corpus and never falling below half in any single year. Long time frames is the second most frequent (63\% overall), with a growing trend in recent years, reaching 86\% in 2025. \response{The figure also shows that papers questioning the growth narrative has the sharpest growing trend (pun intended): from near absence in 2019–2020 (0\% and 7\%) the topic is mentioned in over half of papers in 2024–2025 (50\% and 57\%), suggesting that degrowth and post-growth framing has moved from a marginal to a mainstream concern within the community.}

\subsubsection{Efficiency}

\begin{figure}[h]
    \centering
    \includegraphics[width=1\linewidth]{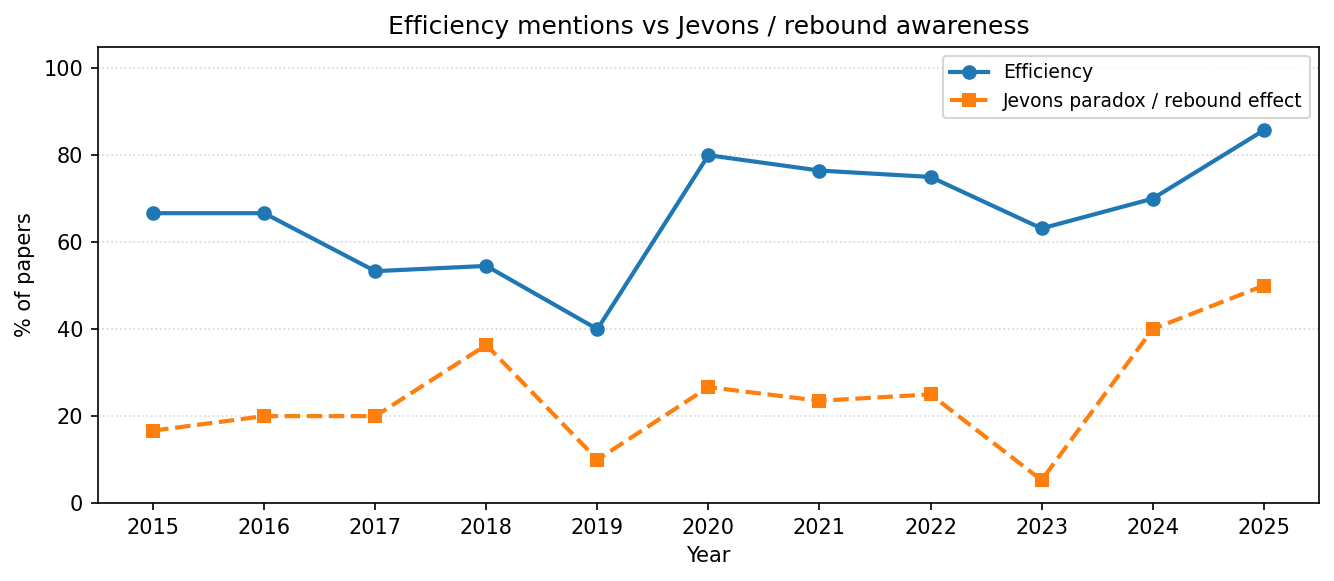}
    \caption{Percentage of articles per year that mention efficiency and, as a subset, those that also engage with the Jevons paradox or rebound effect.}
    \label{fig:efficiency_jevons}
    \Description{}
\end{figure}

Figure~\ref{fig:efficiency_jevons} shows a persistent gap between efficiency language and engagement with the Jevons paradox or rebound effect. Across the time, around 60\% of papers mention efficiency while around 20\% also mention the Jevons paradox or rebound effect. Although both have growing trends in recent years with a smaller gap between the two.

\subsubsection{Global outreach} 
\begin{figure}[h]
  \centering
  \includegraphics[width=\linewidth]{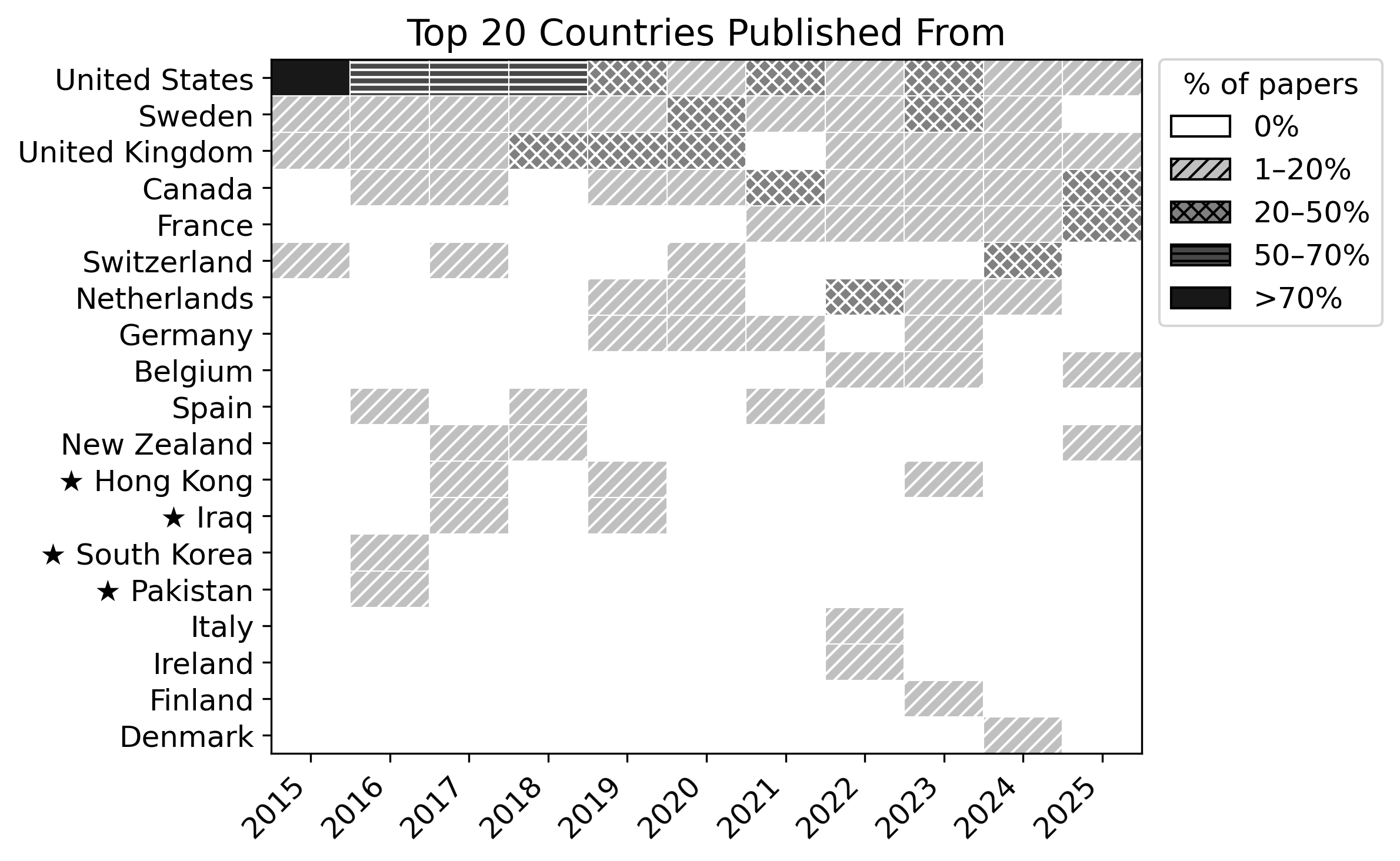}
  \caption{Countries from which papers originate (based on author affiliations), expressed as a percentage of papers published that year. $\bigstar$ = non-WEIRD country}
  \Description{}
  \label{fig:papers_by_country_heat}
\end{figure}

The origins of publications are presented in Figure~\ref{fig:papers_by_country_heat}. It shows that it started US focused and spread to European countries over the years. In the top 20 countries the papers are published from, only four non-WEIRD countries were identified, none of which have published in the recent years.

\begin{figure}[h]
    \centering
    \includegraphics[width=1\linewidth]{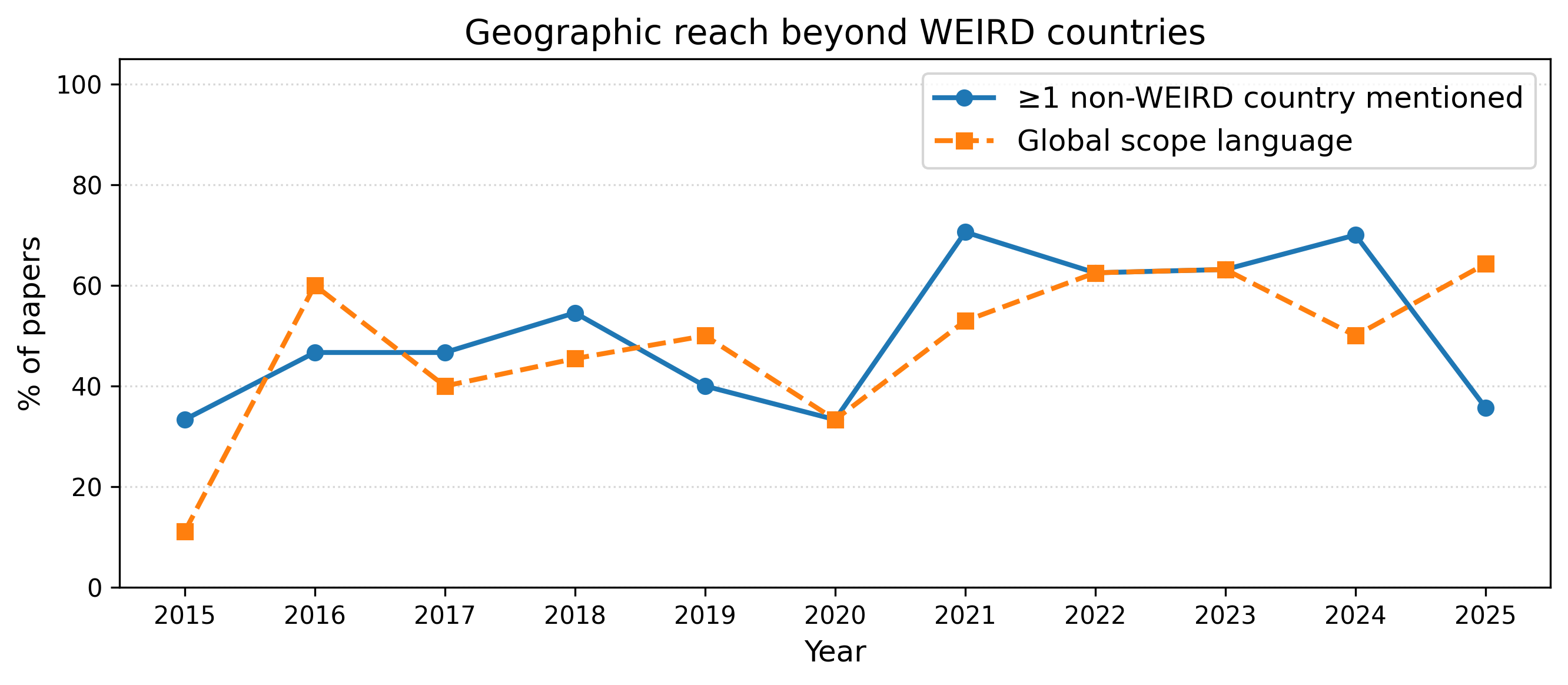}
    \caption{Percentage of the papers per year engaging with geographies beyond WEIRD countries, measured through two signals: non-WEIRD country mentioned and global scope language. }
    \label{fig:nonweird_reach}
    \Description{}
\end{figure}

To see if the content of the publications keeps to the global reach, we can look at Figure~\ref{fig:nonweird_reach}. Overall it seems that around half of the papers have both either mentioned a non-WEIRD country, meaning at least one non-WEIRD country is named in the paper body (e.g. China, India, Kenya). Or the paper used terms indicating a global focus, such as "Global South", "majority world", "developing countries", "low- and middle-income countries" (LMICs), or "cross-cultural". In recent years, after a dip in 2020, the trends have increased to around 60\% of the papers. There has been a decrease in 2025 of mentioning specific non-WEIRD countries, while the global scope language stayed in an upwards trend.

\subsection{Disciplines} 

\begin{figure}[h]
  \centering
  \includegraphics[width=\linewidth]{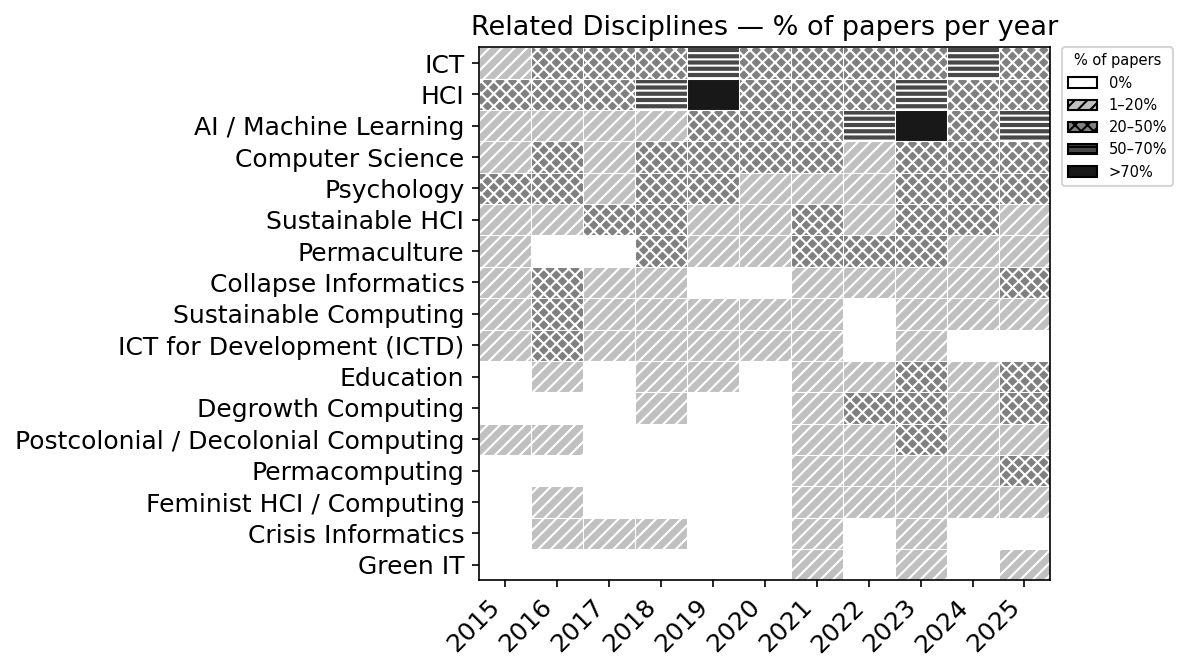}
  \caption{Presence of related disciplines in the papers, expressed as the percentage of papers per year that reference each field. A paper is counted for a discipline if its body text contains characteristic terminology (e.g. "collapse informatics", "decolonial computing", "large language model"). Disciplines are ordered by total frequency across the corpus. Single paper may reference multiple disciplines.}
  \Description{}
  \label{fig:disciplines_heatmap}
\end{figure}

Figure~\ref{fig:disciplines_heatmap} shows that LIMITS papers draw predominantly from HCI and ICT throughout the entire 2015–2025 period, confirming the origin of the field. There is a trend of AI and Machine Learning references from 2022 onwards, going above 70\% in 2023, reflecting community's engagement with the generative AI. Psychology appears as a consistently referenced discipline across the corpus, suggesting LIMITS authors frequently engage with behavioural and cognitive dimensions of computing limits. There was a change from publication from 2021 in discussing topics such as degrowth computing, permacomputing, feminist HCI, and Green IT. 

From our manual review we saw that even though majority papers relate to either computer science or HCI, it has been applied to a variety of problems ranging from education, energy, agriculture to story telling, conservation technologies and music. 

\subsection{\response{LIMITS' Call for Papers through the years}}
\label{sec:cfp}

\response{The LIMITS workshop's "Call for Papers" lays the foundation for the type of submissions the workshop receives, and in this section we will present a short description of how the call has changed over time. We were unable to retrieve the Call for Papers from 2015 and 2016; these two years are therefore omitted here.}

\response{From 2017-2019, the Call for Papers stayed nearly identical, with the aim to "foster research on the impact of present or future ecological, material, energetic, and/or societal limits on computing and computing research to respond to such limits" \cite{limits2017, limits2018, limits2019}. The organizers outlined two broad categories of papers, "discussion papers" and "system papers", without any requirement that papers fit exactly into either category.}

\response{In 2020, the Call for Papers took on a different form, with a particular aim to "reach computing researchers outside of the LIMITS community who might ask: 'what does research in a LIMITS future look like?'" \cite{limits2020}. The call specifically asked for papers on "future systems", while discouraging papers of the category "critical", "analysis" or similar.}

\response{The call in 2021 \cite{limits2021} was similarly framed around a specific question: "What is a LIMITS-aligned computing system?", with an encouragement to submit papers about "hypothetical systems" or "transitional systems". From 2021 and onward, all calls included a note to authors "to consider the stories they tell and reify through their work", with a reminder that "stories have power" \cite{costanza2020design}.}

\response{The 2022 Call for Papers had a more general invitation for works related to "(re)designing computing systems that engage pressing ecological issues", in addition to papers specifically dealing with "transitional systems"~\cite{limits2022}. From this year and forward, the call explicitly included artists and designers in its target group. In 2023, the general invitation included both "ecological and \textit{social} issues", with an additional encouragement for papers building on previous LIMITS papers~\cite{limits2023}.}

\response{The 2024 and 2025 versions were similar to that of 2023, but included a specific invitation for papers on "hypothetical systems" \cite{limits2024, limits2025}. The 2025 version broadened the scope of the general invitation to include both "ecological and social issues and \textit{crises}".}

\response{The various versions of the Call for Papers are generally quite broad, while still having notable differences and adjustments. Given that our review only scratch the surface when it comes to describing the diversity of topics and methodologies employed in the papers, it is challenging to establish correlations between the Calls and the accepted papers. It is nevertheless interesting to see how the workshop's scope has evolved through the years.}


\section{Discussion}

As our aim was to investigate how LIMITS has evolved as a field, we discuss our findings through the lens of the principles defined by Nardi et. al \cite{nardi2018computing} and other notable trends we have identified.

\subsection{Questioning the growth narrative}

LIMITS has engaged with the growth narrative from the start, with many early papers taking a positional or observational stance. Explicit mention of growth critique has grown sharply in recent years, and disciplines that specialise in questioning growth, such as degrowth computing and permacomputing, are increasingly referenced. That said, having the word "limits" in the name naturally draws papers on all kinds of limitations, whether the limits of wireless networks \cite{schmitt2017low} or the limits of human imagination \cite{tanenbaum2016limits}. 

One persistent gap is the disconnect between efficiency language and awareness of the Jevons paradox. Around 60\% of papers mention efficiency, but only around 20\% also engage with the Jevons paradox or rebound effects. Discussing efficiency without acknowledging rebound risks reinforcing the growth mindset the field set out to challenge. At the same time, simply naming the problem is not enough either. 

Following \cite{silberman2023information}, who argues that environmental problems are political rather than purely technical, it makes sense that the field continues to produce positional and observational papers alongside solutions. The solutions that do appear are often not just software or hardware, but guidelines, frameworks and protocols, which reflects a community that sees conceptual work as part of the contribution.

\subsection{Resource scarcity and material reduction}

Scarcity and resource limits is the most consistently present principle across the publications. Concrete artifact production, however, remains low throughout the decade, suggesting the field talks about material reduction more than it builds for it. The artifacts that do appear are often conceptual tools rather than deployed systems. Hardware and software contributions are a smaller group, and they tend to cluster around specific themes such as solar-powered infrastructure \cite{brain2022solar} and device reuse \cite{franquesa2018devices}. 

\response{The low number of artifact production may be because the concept of \textit{computing within limits} points towards less technology rather than more, and that adapting to scarcity and resource limits is more about changing our perception and breaking out of certain narratives, rather than creating new tools.} One of the early LIMITS paper, "Foster the 'mores', counter the 'limits'" \cite{gui2015foster}, makes the point that climate narratives built around having less of everything are hard to argue for politically, and that framing matters. This is where interdisciplinary applications come in use\response{, f}or example computing ideas of refactoring to minimise the systematic complexity within society \cite{raghavan2016refactoring}.

\subsection{Global spatial scale}

The founding workshops drew participants from a range of countries, hosted in the US and \cite{nardi2018computing} frames the field's concerns as global from the outset. Our data shows that authorship has shifted from the US toward Europe, but authorship from other countries than that remains small and has not grown much in recent years. Content engagement is more encouraging: around 60\% of recent papers mention non-WEIRD countries or use global scope language. But writing about a place is not the same as working with people from it, the limitation of our work also does not tell how much the research has engaged with the countries.

The principle "nothing about us without us" applies here. 
Researchers from non-WEIRD countries participated in early workshops but this has not translated into sustained publication. It is worth asking why. \cite{mensah2025missing} argues that technological development is missing African values, and if LIMITS is to genuinely operate at a global scale, it needs to actively work toward including perspectives from beyond Europe and North America, not just reference them. 

\subsection{Permacomputing and permaculture}

One interesting finding is how much permaculture as a framework has been mentioned \response{in} the field. This is perhaps less surprising when you consider that agriculture, food and land care sit naturally alongside scarcity and long-term thinking. Permaculture principles have been discussed and applied in a number of papers \cite{liu2018out, egan2019lions, norton2019sage, egan2020developing, hansen2020app, nelson2024symbiotic}
, and permacomputing, which first appeared in the literature around 2018
, began appearing regularly in LIMITS publications from 2021. This is a shift that \cite{hendry2021transition}, writing in 2021, did not identify as a trend, which suggests it has accelerated since.

\subsection{Disciplines and future directions}

Psychology is consistently referenced across the corpus, and a small but steady share of papers study communities, users or human behaviour. Nilsson et al. \cite{nilsson2024what} also argue for more-than-human approaches, pushing back against the assumption that computing exists purely in service of human needs. This is essential as to answer to Nardi et al. \cite{nardi2018computing} raised question of "How can we maintain or increase well-being while staying within ecological limits?".

Since 2022, AI has appeared as a major concern in the literature \cite{jskelinen2022environmental, beignon2025imposing}, which mirrors broader trends outside the field. AI is relevant to LIMITS in two ways: as a new resource limit to take a position on, given its \response{enormous energy and data center demands \cite{jegham2025hungry}}, and as a tool the community itself can use. \response{The new Green AI field \cite{schwartz2020green} seeks to decrease AI's exponentially growing environmental footprint, but suffers from a narrow focus that neglects Jevons paradox and rebound effects in general \cite{luccioni2025efficiency}. The LIMITS perspective is therefore promising when it comes to imagining possible futures where AI is used (or not used) in a way that is compatible with life within limits.}

\subsection{Limitations}
Constructing the questions and categories all shape the findings before any counting begins. For example the WEIRD framing, borrowed from cross-cultural psychology \cite{henrichWeirdestPeopleWorld2010}, is not a neutral descriptive category but a political one as it names a structural asymmetry that the LIMITS community has itself debated. \response{Classification also carries a degree of uncertainty as countries shift over time in terms of wealth and democratic governance.} In addition, institutional affiliation as reported by metadata does not capture researcher origin or lived experience; a scholar from the Global South affiliated with a European institution appears in our data as European. Country mentions in the papers indicate only that a place was named, not that the study was conducted there or that the researchers had a meaningful connection to it. 

Counting mentions of words also has clear limits. A paper that mentions "scarcity" once is treated the same as one where scarcity is the central concern.


\section{Conclusion}


After a decade of research, the LIMITS community has broadly 
followed through on the principles set out by 
\cite{nardi2018computing}. Growth critique mention has found its way to majority of publications, scarcity and resource limits are consistently brought up across, and the field is beginning to close the gap between efficiency language and awareness of rebound effects. Publication has spread beyond the US to more countries, and at least half of papers each year reference geographies or concerns beyond them, though authorship remains predominantly European and North American. The field is also maturing in its output: solution-oriented papers have grown in recent years, bringing with them hardware and software contributions alongside frameworks, guidelines and design approaches.

LIMITS continues to draw from a wide disciplinary base. HCI and computer science remain central, while newer influences such as degrowth, permacomputing, postcolonial computing and feminist HCI have grown since 2021. The \response{acceleration of AI's resource footprint} since 2022 presents both a challenge and an opportunity. Applying LIMITS principles to AI development is one of the more pressing tasks ahead, \response{countering the trend of chasing efficiency alone.}

The field's interdisciplinary character is one of its strengths. Wicked problems do not respect disciplinary boundaries, and seeing computing ideas applied to agriculture, education, conservation, storytelling and music shows promise of novel ideas. Understanding technology as simultaneously a problem and a resource, considered in full political, material and human context, remains the central contribution of computing within limits.


\subsection{Future work}
In this work we focused on comparing how LIMITS have evolved in comparison to Nardi et al. \cite{nardi2018computing} principles, using a 
combination of programmatic and manual analysis. A complementary 
approach would be a bottom-up topic analysis, identifying what topics arise from the community. In addition, given that several adjacent 
disciplines appear in the LIMITS publications, it would also be worth 
investigating whether computing within limits ideas have 
travelled in the other direction: a systematic review of how 
LIMITS concepts appear in mainstream HCI, ICT for 
sustainability, or degrowth literature would help establish 
whether the field is shaping broader conversations or remaining 
self-contained.


\bibliographystyle{ACM-Reference-Format}
\bibliography{sample-base}

\section*{Papers included in the review}
{\small\noindent\textit{Note: Out of 162 papers published in the LIMITS workshop, we included 160 papers in our review. Two papers were excluded because the full papers could not be found online.}\par\medskip}
\bibliographystylereview{ACM-Reference-Format}
\nocitereview{*}
\vspace{-8mm}
\bibliographyreview{cwl-review-filtered}

\end{document}